\documentclass[a4paper,11pt]{article}

\usepackage{amssymb, amsmath, amsthm, multirow, bbm, url, hyperref}
\usepackage{tikz}
\usepackage[caption=false]{subfig}
\usepackage[margin=2cm]{geometry}

\theoremstyle{plain}

\newtheorem{corollary}{Corollary}
\newtheorem{lemma}{Lemma}
\newtheorem{proposition}{Proposition}
\newtheorem{theorem}{Theorem}
\newtheorem*{claim*}{Claim}

\theoremstyle{definition}
\newtheorem{definition}{Definition}

\newtheorem{example}{Example}

\newcommand{\GF}{\mathrm{GF}}

\newcommand{\Sym}{\mathrm{Sym}}

\newcommand{\GL}{\mathrm{GL}}
\newcommand{\Aff}{\mathrm{Aff}}
\newcommand{\Cay}{\mathrm{Cay}}

\begin{document}

\title{Memoryless computation: new results, constructions, and extensions}

%
%

\author{Maximilien Gadouleau\footnote{School of Engineering and Computing Sciences, Durham University, Durham, UK. Email: \texttt{m.r.gadouleau@durham.ac.uk}} \and S\o{}ren Riis\footnote{School of Electronic Engineering and Computer Science, Queen Mary, University of London, Mile End Road, London E1 4NS, UK. Email: \texttt{smriis@eecs.qmul.ac.uk}}}

\maketitle

\begin{abstract}
In this paper, we are interested in memoryless computation, a modern paradigm to compute functions which generalises the famous XOR swap algorithm to exchange the contents of two variables without using a buffer. This uses a combinatorial framework for procedural programming languages, where programs are only allowed to update one variable at a time. We first consider programs which do not have any memory. We prove that any function of $n$ variables can be computed this way in only $4n-3$ variable updates. We then derive the exact number of instructions required to compute any manipulation of variables. This shows that combining variables, instead of simply moving them around, not only allows for memoryless programs, but also yields shorter programs. Second, we show that allowing programs to use memory is also incorporated in the memoryless computation framework. We then quantify the gains obtained by using memory: this leads to shorter programs and allows us to use only binary instructions, which is not sufficient in general when no memory is used.
\end{abstract}

%

\section{Introduction} \label{sec:intro}

How do you swap the contents of variables $x$ and $y$ using a procedural programming language? The common approach is to use a buffer $t$, and to do as follows (using pseudo-code).

\begin{align*}
    t &\gets x\\
    x &\gets y\\
    y &\gets t.
\end{align*}

However, a famous programmer's trick consists in using XOR, which can be viewed as addition over a binary vector space:
\begin{align*}
    x &\gets x+y\\
    y &\gets x+y\\
    x &\gets x+y.
\end{align*}
The swap can thus be performed without any use of memory. The aim is to generalise this idea to compute any possible function without memory.

Memoryless computation (MC)--referred to as {\em in situ} programs in \cite{Bur96} or computation with no memory in \cite{BGT13}--is a modern paradigm for computing functions, which offers two main innovations. The first introduces a completely different view on how to compute functions. The basic example is the XOR swap described above. Unlike traditional computing, which views the registers as ``black boxes,'' MC takes advantage of the nature of the information contained in those registers and combines the values of the different registers. Thus, it can be seen as the computing analogue of network coding, a revolutionary technique to transmit data through a network which lets the intermediate nodes combine the messages they receive \cite{YLCZ06}. In particular, the XOR swap is the analogue of the canonical example of network coding, the so-called butterfly network \cite{ACLY00}.

The second main innovation lies in the computational model used for MC, which can be briefly described as follows. A processing unit has $n$ registers $x_1,\ldots,x_n$ containing data over a finite alphabet $A$ and has to compute a function $f: A^n \to A^n$ which possibly modifies the values of all registers. It is allowed any updates which only modify one register at a time (i.e., $x_i \gets g(x_1,\ldots,x_n)$ for some $g:A^n \to A$), which are called {\em instructions}. A sequence of instructions computing a given function is a {\em program} for that function. This model aims at emulating computations as they are carried in a core, for they mostly involve manipulations of registers \cite{HP11}. Because an instruction is viewed as a quantum of complexity (similar to a clock cycle), the (procedural) {\em complexity} of a function is defined as the minimum length of a program computing that function. For instance, the complexity of the swap of two bits is equal to three instructions. 

MC has a wide range of possible applications, especially for computationally expensive problems. In this paper, we show how it offers several advantages over traditional computing. First, MC offers a computational speed-up at the core level. Indeed, MC yields arbitrarily shorter programs than traditional computing when manipulating variables (a manipulation of variables is any function of the form $f(x_1,\ldots,x_n) = (x_{1\phi},\ldots,x_{n\phi})$ for some $\phi : \{1,\ldots,n\} \to \{1,\ldots,n\}$) (see Corollary \ref{cor:phi}). Secondly, MC does not rely on additional buffers and hence performs computations in line. Memory management is a tedious task which can significantly slow down computations \cite{HP11}. Although it can be alleviated by using different levels of cache, it still uses more hardware and brings a significant overhead. This problem is particularly important for parallel architectures with shared memory \cite{HP11}. MC offers a radical alternative: it uses no memory at all. It thus eases concurrent execution of different tasks by preventing memory conflicts. It also optimises the use of a crucial and expensive resource and offers another speed-up by avoiding any communication with the data memory. 

While the XOR example described above is folklore, the idea to compute functions without memory was developed  in \cite{Bur96, Bur04, BM00, BM04a, BM04, BGT09}. It is notably proved that any function can be computed without memory. Moreover, only $2n-1$ instructions are needed to compute any bijective function $f: A^n \to A^n$; any function $f: \{0,1\}^n \to \{0,1\}^n$ can be computed in only $4n-3$ instructions. For a complete survey of these results, see \cite{BGT13}.

We would like to emphasize the novelty of the results of this paper and how they differ from those in the literature.
\begin{itemize}
	\item Many aspects considered in this paper are completely novel. These include the study of the average procedural complexity in Sections \ref{sec:further_permutations} and \ref{sec:further_linear}, the study of manipulations of variables in Section \ref{sec:manipulation}, the use of binary instructions in Theorem \ref{th:binary_instructions} and the use of additional registers in Section \ref{sec:memory}.

	\item Some of the results presented in this paper generalise some of those given in the literature. For instance, while \cite{BGT09} proves that any boolean function can be computed in $4n-3$ instructions, we can extend this result to functions over any alphabet. This extension was independently derived in \cite{BGT13}, using a different proof. Other results provide some matching upper and lower bounds which are absent in the literature, e.g. in Theorem \ref{th:2n-1_permutation}. Finally, we also provide alternative proofs to known results. Notably, the proof of the seminal Theorem \ref{th:all} is much more concise; the proof of Theorem \ref{th:2n-1_permutation} highlights some connections with other branches of combinatorics.
\end{itemize}

The rest of the paper is organised as follows. Section \ref{sec:model} reviews the memoryless computation model and proves that it is universal: any function can be computed without memory. Section \ref{sec:complexity} then investigates the number of updates required to compute any function. Section \ref{sec:manipulation} determines the complexity of manipulating variables without memory and shows that memoryless computation yields shorter programs than traditional methods. Section \ref{sec:memory} finally proves that additional registers (or memory) can be added into the memoryless computation model without loss of generality.

\section{Model for memoryless computation} \label{sec:model}

\subsection{Instructions and programs}

We first review the model for memoryless computation introduced in \cite{Bur96} and subsequently developed in \cite{Bur04, BM00, BM04a, BM04, BGT09}.

Let $A$ be a finite set, referred to as the {\em alphabet}, of cardinality $q$ and let $n$ be a positive integer (without loss, we shall usually regard $A$ as $\mathbb{Z}_{q}$ or $\GF(q)$ when $q$ is a prime power). We refer to any element of $A^n$ as a {\em state}. We view any transformation $f$ of $A^n$ (i.e., $f: A^n \to A^n$) as a tuple of functions $f = (f_1,\ldots,f_n)$, where $f_i : A^n \to A$ is referred to as the $i$-th coordinate function of $f$. In particular, a coordinate function is {\em trivial} if it is equal to the identity, i.e. $f_i(x) = x_i$; it is nontrivial otherwise. The size of the image of $f$ is referred to as its {\em rank}. When considering a sequence of transformations, we shall use superscripts, e.g. $f^k: A^n \to A^n$ for all $k$--and hence $f^k$ shall never mean taking $f$ to the power $k$.

\begin{definition}[Instruction] \label{def:instruction}
An {\em instruction} is a transformation $g$ of $A^n$ with at most one nontrivial coordinate function $g_i$. We say that the instruction {\em updates} $y_i$ for $y = (y_1,\ldots,y_n) \in A^n$ and we denote it as
$$
    y_i \gets g_i(y).
$$
A permutation instruction is an instruction which maps $A^n$ bijectively onto $A^n$ (i.e. is a permutation of $A^n$).
\end{definition}

By definition, the identity is an instruction, which can be represented by
$y_i \gets y_i$ for any $1 \le i \le n$.

We denote the set of instructions of $A^n$ as $\bar{\mathcal{I}}(A^n)$ and the set of permutation instructions as $\mathcal{I}(A^n)$. We shall simply write $\bar{\mathcal{I}}$ and $\mathcal{I}$ when there is no ambiguity. For instance, if $A = \GF(2)$ and $n=2$, then $\mathcal{I}$ is given by
$$
    \{ (x_1,x_2), (x_1+1,x_2), (x_1+x_2,x_2), (x_1+x_2+1,x_2), (x_1,x_2+1), (x_1,x_1+x_2), (x_1,x_1+x_2+1) \}.
$$
In update form, $\mathcal{I}$ can be written as follows:

\begin{table}[!h]
\begin{center}
\begin{tabular}{llll}
    $\{ y_1 \gets y_1$,& $y_1 \gets y_1 + 1$,& $y_1 \gets y_1 + y_2$,& $y_1 \gets y_1 + y_2 +1$,\\
     $y_2 \gets y_2$,& $y_2 \gets y_2 + 1$,& $y_2 \gets y_1 + y_2$,& $y_2 \gets y_1 + y_2 +1\}$,
\end{tabular}
\end{center}
\end{table}

\noindent where the identity is represented by $y_1 \gets y_1$ and $y_2 \gets y_2$.

\begin{definition}[Program]
For any transformation $f$ of $A^n$, a {\em program} of length $L$ computing $f$ is a sequence of instructions $g^1,\ldots,g^L$ such that
$$
    f = g^L \circ \ldots \circ g^1.
$$
\end{definition}

We shall write the instructions of a program in their update form one below the other. Although the identity is an instruction, any instruction in a program is not the identity unless specified otherwise. Also, since the set of instructions updating a given coordinate is closed under composition, without loss we can always assume that $g^{k+1}$ updates a different coordinate than $g^k$ for all $k$. The cases where $q = 1$ or $n=1$ being trivial, we shall assume $q \ge 2$ and $n \ge 2$ henceforth.

We consider a processor core which has access to a finite number of registers and only allows programs of the form described above. Therefore, it only allows in-place calculations, without loops, pointers, and more importantly without any memory. We use $y = (y_1,\ldots,y_n)$ to represent the content of the registers during the program, $x$ to represent the input and $f(x)$ to represent the output. Hence $y = x$ before the first instruction, and $y = f(x)$ after the last instruction. Note that we will also use the shorthand notation $y_i \gets h(x)$ to reflect how the content of the registers relates with the program input. In particular, note that the last update of $y_i$ must be
$$
    y_i \gets f_i(x).
$$
To be absolutely rigorous, we should let $y$ take into account the instruction number: $y^{0} = x, y^{1}, \ldots, y^{L} = f(x)$, where $L$ is the length of the program. However, our calculations will not require such level of rigour, and we simply use $y$ instead.

In order to illustrate the notation, let us rewrite the program computing the swap of two variables, i.e. $f: A^2 \to A^2$ where $f(x_1,x_2) = (x_2,x_1)$. It is given as follows (all operations being done mod $q$):
\begin{align*}
    y_1 &\gets y_1 + y_2 \qquad (= x_1 + x_2)\\
    y_2 &\gets y_1 - y_2 \qquad (= x_1)\\
    y_1 &\gets y_1 - y_2 \qquad (= x_2).
\end{align*}

\begin{definition}
Let $B,C$ be two alphabets and $f,g: B \to C$. We say $g$ {\em dominates} $f$ if $g(x) = g(x') \Rightarrow f(x) = f(x')$
for all $x, x' \in B$. In other words, $f = h \circ g$ for some transformation $h$.
\end{definition}

A program for $f$ induces a sequence of transformations $h^1,\ldots,h^L = f$ where $h^1$ is an instruction, $h^i$ and $h^{i+1}$ differ in only one coordinate, and $h^i$ dominates $h^{i+1}$ for all $i$. Indeed, simply let $h^{i+1} = g^{i+1} \circ h^i$; equivalently $h^i$ represents the content of $y$ after the $i$-th instruction of the program. In particular, if $f$ is a permutation, then all intermediate transformations must be permutations as well.

We remark that this framework only allows to return one output: the transformation $f$ computed by the program. However, it may be fair to ask the program to sequentially return outputs. This can be incorporated in this framework if all the outputs are permutations. However, the case of general transformations is more troublesome: for instance, if we ask to return $f^1(x_1,x_2) = (x_1,x_1+1)$ and then $f^2(x_1,x_2) = (x_2,x_2+1)$, then it is clear that $f^2$ cannot be computed after $f^1$. In general, a program can sequentially compute $f^1, \ldots, f^K$ only if $f^i$ dominates $f^{i+1}$ for all $1 \le i \le K-1$ (the results in this paper will show that this is necessary and sufficient). Therefore, this program can be broken down into $K$ shorter programs, each computing one output. In view of these considerations, we shall only consider programs which compute one output transformation $f$ in the remaining of this paper.

\subsection{All transformations are computable without memory}

We are now interested in the general case of computing any transformation of $n$ variables. We first give in Theorem \ref{th:all} a new, more concise, proof that any transformation can be computed without memory, a result from \cite{Bur96}. Although the program in the proof has exponential length, we shall see that any transformation has a program of linear length in Section \ref{sec:complexity}.

We introduce some useful notations for any states $u,v \in A^n$. First, the {\em transposition} of $u$ and $v$, denoted as $(u, v)$, is the permutation of $A^n$ which maps $u$ to $v$, $v$ to $u$, and fixes any other state in $A^n$. Second, the {\em assignment} of $u$ to $v$, denoted as $(u \to v)$, is the transformation which maps $u$ to $v$ and fixes any other state in $A^n$. Third, we denote the all-zero state as $e^0$ and the $k$-th unit state as $e^k \in A^n$, where $e^k_i = \delta(i,k)$ and $\delta$ is the Kronecker delta function. Therefore, if $v = u + e^i$ for some $i$, the transposition $(u,v)$ is an instruction with update form
$$
    y_i \gets y_i + \delta(y,u) - \delta(y,v),
$$
Moreover, the assignment $(e^0 \to e^1)$ is an instruction with update form
$$
	y_1 \gets y_1 + \delta(y,e^0).
$$

\begin{theorem} \label{th:all}
Any transformation of $A^n$ can be computed by a program which only consists of transpositions $(u, v)$ where $v = u + e^i$ for some $i$ and the assignment $(e^0 \to e^1)$. 
\end{theorem}

\begin{proof}
If we order the states of $A^n$ according to the Gray code in \cite{Gua98}, then any two consecutive states $v^j$ and $v^{j+1}$ satisfy $v^{j+1} = v^j \pm e^{i_j}$ for some $i_j$. The transpositions above are exactly the Coxeter generators $\{(v^j,v^{j+1}) : 1 \le j \le q^n-1\}$ corresponding to this ordering. Therefore, any permutation of $A^n$ can be computed using these instructions. Furthermore, adding any transformation of rank $q^n-1$ to a generating set of $\Sym(A^n)$ yields a generating set of the transformation monoid of $A^n$ \cite[Theorem 3.1.3]{GM09}. Since the assignment $(e^0 \to e^1)$ is an instruction of rank $q^n-1$, we obtain the result.
\end{proof}

\section{Procedural complexity} \label{sec:complexity}

\begin{definition}[Procedural complexity]
The shortest length of a program computing $f$ is referred to as the {\em procedural complexity} of $f$ and is denoted as $\mathcal{L}(f)$. By convention, the identity has procedural complexity $0$.
\end{definition}

We have $\mathcal{L}(f \circ g) \le \mathcal{L}(f) + \mathcal{L}(g)$ for any two transformations $f$ and $g$. Furthermore, if $f$ is a permutation, then it is easy to show that $\mathcal{L}(f^{-1}) = \mathcal{L}(f)$. We then obtain that
$$
    d(f,g) := \mathcal{L}(f \circ g^{-1})
$$
defines a metric on the symmetric group of $A^n$. This is indeed the word metric, with generators given by all the permutation instructions.

We would like to emphasize that the procedural complexity strongly differs from other measures seen in complexity theory. For instance, the procedural complexity of any decision problem is simply $1$, for it can be expressed as computing the instruction whose value is $1$ if the instance has an affirmative answer and $0$ otherwise. Also, the procedural complexity is based on the set of all instructions, and not only on circuits formed of certain types of gates, unlike in circuit complexity. Therefore, each instruction can be arbitrarily ``complex.''

We believe that the model for memoryless computation is appropriate to evaluate the true complexity of computations operated on cores. Indeed, such computations mostly involve manipulations of registers. Also, the only accurate measure of complexity would be the time it takes for a processor to perform that computation. Because each instruction is counted equal, regardless its nature, the procedural complexity model only takes the number of clock cycles it takes to compute a given function. Obviously, the model remains theoretical for it allows any possible update; the search for efficient instruction sets is work in progress.

\subsection{Procedural complexity of permutations}

The main purpose of this section is to prove that the maximum procedural complexity of a permutation in $\Sym(A^n)$ is $2n-1$, which is independent from the cardinality of the alphabet $A$. The $2n-1$ upper bound was already given in \cite{BGT09}, here we give the matching lower bound and a slightly different proof of the upper bound which highlights its relation to the study of coordinate functions and the so-called combinatorial representations introduced in \cite{CGR13}.

Proposition~\ref{prop:ab} below shows that this quantity is at least $2n-1$. It is remarkable that the permutation which maximises the procedural complexity is very ``simple'' to describe; this fact highlights the difference between the procedural complexity and other complexity measures.

\begin{proposition} \label{prop:ab}
The procedural complexity of the transposition $(a, b)$ of two states $a,b \in A^n$ is $2d-1$ instructions, where $d$ is the Hamming distance between $a$ and $b$:
$d = |\{i: a_i \neq b_i\}|.$
\end{proposition}

\begin{proof}
Without loss, let $a$ and $b$ disagree on their $d$ first coordinates. Denoting 
$$
	v^k = (b_1,\ldots,b_k,a_{k+1},\ldots,a_n)
$$ 
for $1 \le k \le d$, we obtain
$$
    (a, b) = (a, v^1)\circ \cdots\circ (v^{d-2}, v^{d-1}) \circ (v^{d-1}, b)\circ \cdots \circ (v^1, v^2)\circ (a, v^1).
$$
Each transposition involves states differing in at most one position, and hence is an instruction. For instance, $(a, v^1)$ is the instruction
$$
    y_1 \gets y_1 + (b_1-a_1)\left(\delta(y,a) - \delta(y, v^1)\right).
$$
Therefore, the procedural complexity is at most $2d-1$ instructions.

Conversely, suppose that there exists a program computing $(a, b)$ with fewer than $2d-1$ instructions. In that program, at least two coordinates are only updated once (say $i$ before $j$). Denote the images of $a$ and $b$ before the update of $y_j$ as $a'$ and $b'$, respectively. Note that $a'_i = b_i$ and $b'_i = a_i$, since $y_i$ will not be updated any further. The update of $y_j$ is given by
$$
    y_j \gets y_j + (b_j - a_j)(\delta(y,a') - \delta(y,b')),
$$
since coordinate $j$ cannot be modified for any program input other than $a$ or $b$, and it must indeed give the correct values for these two inputs. However, this update is not bijective, for $a'$ and $b'$ differ in coordinate $i$.
\end{proof}

To prove an upper bound on the procedural complexity, we need to study the properties of functions. We use the terminology of \cite{CGR13}. Although this upper bound was proved in \cite{BGT09}, we give an alternate proof below, which connects the topic of this paper to the study of coordinate functions and combinatorial representations from \cite{CGR13}.

\begin{definition} \label{def:balanced}
Let $B,C$ be two alphabets. A function $f : B \times C \to B$ is {\em balanced} if $|f^{-1}(b)| = |C|$ for all $b \in B$.
\end{definition}

It is easily shown that for any two functions $f: B \times C \to B$ and $h: B \times C \to C$, the function $(f,h): B \times C \to B \times C$ is a permutation of $B \times C$ if and only if $f$ is balanced and $h(f^{-1}(b)) = C$ for all $b \in B$ \cite{CGR13}.

\begin{proposition} \label{prop:exchange}
For any pair of balanced functions $f,g : B \times C \to B$, there exists $h : B \times C \to C$ such that $(f,h)$ and $(g,h)$ are permutations of $B \times C$.
\end{proposition}

\begin{proof}
Let $G$ be the bipartite graph with vertex set given by two copies of $B$ and an edge between $i$ and $j$ for each element of $(f,g)^{-1}(i,j) \subseteq B \times C$. Since $f$ and $g$ are balanced, $G$ is $|C|$-regular and hence its edges are colourable with colours from $C$ (this is an easy consequence of \cite[Corollary 16.6]{BM08}). Let $h: B \times C \to C$ be such a colouring. Then for all $i \in B$, we have $h(f^{-1}(i)) = \bigcup_{j \in B} h((f,g)^{-1}(i,j)) = C$ and similarly $h(g^{-1}(j)) = C$. This is equivalent to $(f,h)$ and $(g,h)$ being permutations.
\end{proof}

\begin{theorem} \label{th:2n-1_permutation}
The maximum procedural complexity of a permutation  of $A^n$ is $2n-1$ instructions.
\end{theorem}

\begin{proof}
By Proposition~\ref{prop:ab}, we only need to prove that any permutation $f$ can be computed by a program with at most $2n-1$ instructions. We prove the following claim: for any $1 \le k \le n-1$, there exists a function $h_k : A^n \to A$ of $x$ such that $(h_1,\ldots,h_k,x_{k+1},\ldots,x_n)$ and $(h_1,\ldots,h_k,f_{k+1},\ldots,f_n)$ are permutations. This is clear for $k=1$: apply Proposition~\ref{prop:exchange} to $(f_2,\ldots,f_n)$ and $(x_2,\ldots,x_n)$. Let us assume it is true for up to $k-1$, then by hypothesis, $g^1 := (h_1,\ldots,h_{k-1},x_{k+1},\ldots,x_n)$ and $g^2 := (h_1,\ldots,h_{k-1},f_{k+1},\ldots,f_n)$ are both balanced functions from $A^n$ to $A^{n-1}$ (since $(g^1,x_k)$ and $(g^2,f_k)$ are permutations, respectively). Applying Proposition~\ref{prop:exchange} to these functions then proves the claim.

The program then proceeds as follows:
\begin{itemize}
    \item Step 1. For $k$ from $1$ to $n-1$, do $y_k \gets h_k(x).$

    \item Step 2. For $k$ from $n$ down to $1$, do $y_k \gets f_k(x).$
\end{itemize}
\end{proof}

\subsection{Further results for permutations} \label{sec:further_permutations}

We can represent computations of any permutation of $A^n$ as progressing around the Cayley graph \cite{GR01} $\Cay(\Sym(A^n), \mathcal{I})$. The set of permutation instructions $\mathcal{I} \subseteq \Sym(A^n)$ is described as follows. Let $g$ be the instruction $y_i \gets g_i(y)$. Then in view of the remarks made after Definition~\ref{def:balanced}, $g$ is a permutation if and only if $g_i: A^n \to A$ satisfies
$$
    g_i(\{u \in A^n : (u_1,\ldots,u_{i-1},u_{i+1},\ldots,u_n) = v\}) = A
$$
for all $v \in A^{n-1}$. There are hence $q!$ choices for the reduction of $g_i$ to each pre-image, and hence $(q!)^{q^{n-1}}$ choices for $g_i$. Since the identity has been counted $n$ times, there are
$$
    |\mathcal{I}| = n(q!)^{q^{n-1}} - (n - 1)
$$
instructions. We remark that the set of permutation instructions updating a given coordinate forms a group, isomorphic to $\Sym(A)^{q^{n-1}}$.


We have determined the maximum procedural complexity in Theorem~\ref{th:2n-1_permutation}. We are now interested in the average complexity. Proposition~\ref{prop:almost_all_permutations} gives a lower bound on that quantity.

\begin{proposition} \label{prop:almost_all_permutations}
The proportion of permutations with computational complexity at least
$$
    \left\lfloor \frac{n \log q - 1}{q^{-1}\log q! + q^{-n}\log n} \right\rfloor + 1
$$
tends to $1$ when $n$ tends to infinity.
\end{proposition}

\begin{proof}
Any transformation with procedural complexity $l$ can be expressed as a product of $l$ instructions. Therefore, the number of permutations with procedural complexity at most $l$ is no more than the number of $l$-tuples of permutation instructions, given by $|\mathcal{I}|^l$. We have
\begin{align*}
    |\mathcal{I}| &\le n(q!)^{q^{n-1}} = \exp(\log n + q^{n-1} \log q!),\\
    |\Sym(A^n)| = q^n! &\ge \sqrt{2 \pi q^n} q^{nq^n} \exp(-q^n) = \sqrt{2 \pi q^n} \exp(q^n(n\log q - 1)).
\end{align*}

Denoting $B = \frac{n \log q - 1}{q^{-1}\log q! + q^{-n}\log n}$ we obtain $|\Sym(A^n)| \ge \sqrt{2 \pi q^n} |\mathcal{I}|^B$ and hence the proportion of permutations with procedural complexity at most $\lfloor B \rfloor$ is upper bounded by $(2\pi q^n)^{-1/2}$.
\end{proof}

In particular, Proposition~\ref{prop:almost_all_permutations} shows that for $n$ large, almost all permutations of $\GF(2)^n$ have computational complexity at least $2n-2$. Therefore, they are very close to the maximum of $2n-1$. However, the bound in Proposition~\ref{prop:almost_all_permutations} decreases with $q$.

We now show how the problem of determining the procedural complexity of a given permutation can be reduced to the case of so-called ordered permutations for nearly all permutations.

\begin{definition}[Ordered function]
Let $A$ and $A^n$ be ordered (say, using the lexicographic order). For any balanced function $f_i: A^n \to A$ and any $a \in A$, we denote the minimum element of $f_i^{-1}(a)$ as $m(a)$. We say $f_i$ is ordered if $m(0) \le m(1) \le \ldots \le m(q-1)$.
\end{definition}

Any function $f_i: A^n \to A$ can be uniquely expressed as $f_i = \sigma_i \circ f_i^*$ where $\sigma_i \in \Sym(A)$ and $f_i^*$ is ordered. In this case, we say that $f_i$ is {\em parallel} to $f_i^*$ \cite{CGR13}.

By extension, we say that $f$ is ordered if all its coordinate functions are ordered. Therefore, to any permutation $f$, we associate the ordered permutation $f^*$ where $f_i = \sigma_i \circ f_i^*$ for some $\sigma_1,\ldots,\sigma_n \in \Sym(A)$.

\begin{proposition} \label{prop:ordered}
There exists a shortest program computing $f^*$ using only ordered instructions. Furthermore, its length satisfies
$$
   \mathcal{L}(f^*) \le \mathcal{L}(f) \le \mathcal{L}(f^*) + T(f),
$$
where $T(f)$ is the number of nearly trivial (parallel to the trivial coordinate function) coordinate functions of $f$:
$$
    T(f) = |\{i: f_i^* = x_i, f_i \ne x_i \}|.
$$
\end{proposition}

\begin{proof}
We first prove that there exists a shortest program computing $f^*$ using only ordered instructions. Let $f^* = g^L \circ \ldots g^1$ be a shortest program computing $f^*$. We can easily convert it to another program $h^L \circ \ldots \circ h^1$ also computing $f^*$ using only ordered instructions as follows. First let $h^1 = g^{1*}$. Then before $g^j$, we can express the content of the $i$-th cell as $y_i = \rho_i \circ y_i^*$ for all $1 \le i \le n$. Replace the instruction $y_i \gets g^j_i(y)$ by
$$
    y_i \gets h^j_i(y) = \tau g^j_i(\rho_1 \circ y_1, \ldots, \rho_n \circ y_n),
$$
where $\tau \in \Sym(A)$ guarantees that the instruction $h^j$ is indeed ordered. It is easy to check that converting all instructions in this fashion does yield a program computing $f^*$.

We now prove that $\mathcal{L}(f^*) \le \mathcal{L}(f)$. Consider a shortest program $g^L \circ \ldots \circ g^1$ computing $f$ and convert it as follows to compute $f^*$. First, replace any final update $y_i \gets f_i(x)$ by $y_i \gets \sigma_i^{-1} \circ f_i(x) = f_i^*(x)$. Second, after this final update, replace any occurrence of $y_i$ by $\sigma_i y_i$.

We finally prove that $\mathcal{L}(f) \le \mathcal{L}(f^*) + T(f)$. Consider a shortest program $h^L \circ \ldots \circ h^1$ computing $f^*$ (note that it may or may not update any of the coordinates $y_i$ for which $f_i$ is nearly trivial) and convert it as follows to compute $f$. First, replace any final update $y_i \gets f^*_i(x)$ by $y_i \gets \sigma_i \circ f_i^*(x) = f_i(x)$. Second, after this final update, replace any occurrence of $y_i$ by $\sigma_i^{-1} y_i$. Third, update the eventual nearly trivial coordinate functions which have not been updated yet (there are at most $T(f)$ of them).
\end{proof}

\subsection{Program computing linear transformations} \label{sec:further_linear}

We are now concerned with the case where $q$ is a prime power and the inputs $x_1,\ldots,x_n$ are elements of a finite field $A = \GF(q)$, and we want to compute a linear transformation $f$ of $A^n$, i.e.
$$
    f(x) = xM^\top
$$
for some matrix $M \in A^{n \times n}$. Each coordinate function $f_i$ of $f$ can be viewed as the inner product of a row of $M$ with the input vector $x$. Therefore, we shall abuse notations slightly and refer to that row as $f_i$: $f_i(x) = f_i \cdot x$. In this section, we restrict ourselves to linear instructions only, i.e. instructions of the form
$$
    y_i \gets a \cdot y = \sum_{j=1}^n a_j y_j,
$$
for some $a = (a_1,\ldots,a_n) \in A^n$.

Computing $f$ is equivalent to calculating the matrix $M$ as a product of matrices $M = M_1 \ldots M_L$, where $M_i$ is a matrix which only modifies one row. If $M$ is nonsingular, this is also equivalent to a sequence of matrices $N_0 = I_n,N_1,\ldots,N_{L-1},N_L = M$ where $N_i$ is nonsingular and $N_i$ and $N_{i+1}$ only differ by one row for all $i$.

Gaussian elimination indicates that any matrix can be computed by linear instructions involving only two rows. The number of such instructions required to compute any matrix is on the order of $n^2$. However, since we allow any linear instruction involving all $n$ rows, we can obtain shorter programs. In \cite{BGT09}, it is proved that all matrices can be computed in $2n-1$ linear instructions; in fact, their result holds not only for finite fields but for a much larger class of rings.

Let us characterise the set $\mathcal{M}(\GF(q)^n)$ of invertible linear instructions.  It is given by the set of nonsingular matrices with at most one nontrivial row:
$\mathcal{M} = \{S(i,v) : 1 \le i \le n, v \in A^n(i)\},$
where
\begin{align*}
    A^n(i) &= \{v \in A^n, v_i \neq 0\} \,\mbox{for all}\, 1 \le i \le n,\\
    S(i,v) &= \left(\begin{array}{c|c|c}
        I_{i-1} & \multicolumn{2}{c}{0}\\
        \hline
        \multicolumn{3}{c}{v}\\
        \hline
        \multicolumn{2}{c|}{0} & I_{n-i}
    \end{array}\right) \in A^{n \times n}.
\end{align*}
Remark that $S(i,v)^{-1} = S(i,-v_i^{-1}v)$ for all $i,v$ and
$|\mathcal{M}| = nq^{n-1}(q-1) - n+1.$
Computing a nonsingular matrix is hence equivalent to progressing around the Cayley graph
$G := \Cay(\GL(n,q),\mathcal{M}).$
Our previous results imply that $G$ is undirected and connected. 
The following are equivalent:
\begin{enumerate}
    \item $M$ and $N$ are adjacent in $G$.

    \item $M = S(i,v)N$ and $N = S(i,-v_i^{-1}v)M$ for some $i$ and $v \in A^n(i)$.

    \item $M$ and $N$ only differ in one row.
\end{enumerate}
Therefore, $G$ is the subgraph of the Hamming graph $H(n,q^n)$ induced by $\GL(n,q)$.

The diameter of $G$ is of great interest as it gives the maximum procedural complexity $\mathcal{L}'(M)$ of computing a nonsingular matrix by updating one row at a time. We know that it is no more than $2n-1$ by \cite{BGT09}; we shall see that it is at least $\lfloor \frac{3n}{2} \rfloor$ (and hence it is equal to $3$ when $n=2$) but it remains unknown for $n \ge 3$. However, when the field $A$ is large, then almost all $n \times n$ matrices can be computed in no more than $n$ linear instructions. The result beow should be compared to the average procedural complexity result of Proposition \ref{prop:almost_all_permutations}.

\begin{proposition} \label{prop:almost_all_matrices}
There are exactly
$$
    (q-1)^nq^{n(n-1)} = \left(1 - \frac{1}{q}\right)^n q^{n^2}
$$
$n \times n$ nonsingular matrices over $\GF(q)$ which can be computed simply by updating their rows from $1$ to $n$ in increasing order.
\end{proposition}

\begin{proof}
Let us count such matrices $M$ with rows $f_i$. After the first instruction, we obtain the matrix whose first row is equal to $f_1$, while the last $n-1$ rows do not depend on the matrix we are computing and are equal to $(0|I_{n-1})$. Then $f_1$ can be chosen as any vector not in the span of the last $n-1$ rows: there are hence $(q-1)q^{n-1}$ choices for $f_1$. Once $f_1$ is fixed, similarly there are $(q-1)q^{n-1}$ choices for $f_2$, and so on.
\end{proof}

Similar to the general case, we can reduce the problem of determining the complexity of nearly any nonsingular matrix to the case of so-called {\em scaled} matrices. Note that this concept is not necessarily consistent with the concept of ordered permutations; however, it can be viewed as an analogue.

\begin{definition}
A nonzero vector whose leading nonzero coefficient is equal to $1$ is said to be {\em scaled}. A nonsingular matrix is {\em scaled} if all its rows are scaled.
\end{definition}

For instance, the identity matrix is the only scaled diagonal matrix. For any nonzero vector $v \in \GF(q)^n$ with leading nonzero coordinate $v_j$, then $v^* := v_j^{-1} v^*$ is a scaled vector. For any nonsingular matrix $M$ with rows $f_i$, let $M^*$ be the corresponding scaled matrix with rows $f_i^*$. We obtain the linear analogue of Proposition~\ref{prop:ordered}.

\begin{proposition} \label{prop:scaled}
There exists a shortest linear program computing $M^*$ with only scaled instructions. Its length satisfies
$$
    \mathcal{L}'(M) \le \mathcal{L}'(M) \le \mathcal{L}'(M^*) + T'(M),
$$
where $T'(M)$ is the number of nearly trivial (equal to multiples of the corresponding unit vectors) rows of $M$:
$$
    T'(M) = |\{i : f_i = \mu_i e^i, \mu_i \in \GF(q) \backslash \{0,1\} \}| = |\{i : f_i \ne e^i, f_i^* = e^i\}|.
$$
\end{proposition}

\subsection{Procedural complexity of all transformations}

We have seen that any permutation of $A^n$ can be computed in $2n-1$ memoryless instructions. We now prove that any transformation can be computed in $4n-3$ memoryless instructions. Although the $2n-1$ bound for permutations is tight, the $4n-3$ bound for general transformations is not: it is easy to check that for $q=2$ and $n=2$, any transformation of $\{0,1\}^2$ can be computed in at most three instructions.

The $4n-3$ bound was already obtained in \cite{BGT09} for the binary alphabet. Although our proof follows a similar structure to the one therein, the key of the extension to any alphabet is Lemma \ref{lem:proper}, which relies on additive number theory. The same extension is also proved in \cite{BGT13}, using a different generalisation to an arbitrary alphabet.

\begin{definition}[Lexicographic order]
For any $a = (a_1,\ldots,a_n) \in A^n$ ($A = \mathbb{Z}_q$), the {\em lexicographic order} of $a$ is the integer $\sum_{i=1}^n a_i q^{i-1}$. For the sake of conciseness, we shall abuse notation and identify $a$ with its lexicographic order.
%
%
An {\em interval} of $A^n$ is any subset of the form
$$
    [b,c) := \{x \in A^n: b \le x < c \}
$$
for any $0 \le b \le c \le q^n-1$. For any $0 \le i \le n$ and $0 \le j < q^{n-i}$, the {\em $j$-th block of level $i$} is the interval
$$
	B_{i,j} := [jq^i, (j+1)q^i).
$$
\end{definition}

We let $\lambda$ be an integer partition of $q^n$, i.e. $\lambda : A^n \to \mathbb{Z}$, where $\lambda_a \ge 0$ for all $a \in A^n$ and $\sum_{a \in A^n} \lambda_a = q^n$. 

\begin{definition}
We say $\lambda$ is {\em proper} if for all $0 \le i \le n$ and all $0 \le j < q^{n-i}$,
$$
	\sum_{a \in B_{i,j}} \lambda_a = 0 \mod q^i.
$$
\end{definition}

\begin{lemma} \label{lem:proper}
Any integer partition $\lambda$ of $q^n$ can be sorted properly, i.e. there exists $h \in \Sym(A^n)$ such that $\lambda \circ h$ is proper.
\end{lemma}

\begin{proof}
We first prove the following claim. Any sequence of $rq$ elements $a_0,\ldots,a_{rq-1}$ of $\mathbb{Z}_q$ satisfying $a_0 + \ldots + a_{rq-1} = 0$ can be re-ordered such that 
$$
	a_0 + \ldots + a_{q-1} = a_q + \ldots + a_{2q-1} = \ldots = a_{(r-1)q} + \ldots + a_{rq-1} = 0.
$$
The proof easily follows from the following theorem due to Erd\"os, Ginzburg and Ziv \cite{EGZ61}: any sequence $b_0,\ldots, b_{2q-2}$ of $2q-1$ elements of $\mathbb{Z}_q$ can be re-ordered such that $b_0 + \ldots + b_{q-1}  = 0$. 

We now build the ordering recursively. Begin at level $i=0$ with $q^n$ blocks of size $1$ having each value in the sequence $\lambda_0,\ldots,\lambda_{q^n-1}$. At level $i+1$, gather the elements of the sequence into groups of $q$ elements, whose values sum up to a multiple of $q$ (this is possible due to our claim), say $kq$. Define the value of this new block as $k$. This defines a new sequence of $q^{n-i-1}$ non-negative integers whose sum is $q^{n-i}$. We finish at level $n$.
\end{proof}

\begin{example}
Example of construction of $\lambda$: let $q = n = 3$ and $\lambda = (5,4,4,3,3,2,2,1,1,1,1,0,\ldots,0)$ with 16 zeros. We obtain the following, where the subscripts denote the value of the block in the next level:
\begin{align*}
	&[5,4,3]_4[4,3,2]_3[2,1,0]_1[1,1,1]_1[0,0,0]_0[0,0,0]_0[0,0,0]_0[0,0,0]_0[0,0,0]_0\\
	&[4,1,1]_2[3,0,0]_1[0,0,0]_0\\
	&[2,1,0]_1
\end{align*}
Therefore, the proper partition is $(5,4,3,2,1,0,1,1,1,4,3,2,0,\ldots,0)$ with 15 zeros at the end.
\end{example}

Now that we have the key arithmetic property of Lemma \ref{lem:proper}, the rest of the proof follows \cite{BGT09}.

\begin{definition}
For any proper integer partition $\lambda$ of $q^n$, $\Lambda$ is the transformation of $A^n$ such that
$$
    \Lambda \left( \left[\sum_{b=0}^{a-1} \lambda_b, \sum_{b=0}^a \lambda_b - 1 \right] \right) = a
$$ 
for all $a \in A^n$.
\end{definition}

\begin{lemma} \label{lemma:compatible_suffix}
For any proper $\lambda$, the transformation $\Lambda$ satisfies the following property:  if $a , b \in A^n$ agree on coordinates $i$ to $n$ for some $i$, then so do $\Lambda(a)$ and $\Lambda(b)$.
\end{lemma}

\begin{proof}
First, we prove the following claim. For every $i,j$ as above, there exist $0 \le k,k' < q^{n-i}$ such that
$\Lambda^{-1}(B_{i,j}) = \bigcup_{k \le l \le k'} B_{i,l}.$

Proof of claim: We remark that the pre-image by $p$ of an interval is an interval itself. By definition of $\Lambda$, we have 
$$
	|\Lambda^{-1}(B_{i,j})| = \sum_{a \in B_{i,j}} \lambda_a = 0 \mod q^i,
$$
since $\lambda$ is proper.

For a fixed $i$ we prove the claim by induction on $j$. If $j=0$ then $|\Lambda^{-1}(B_{i,0})| =kq^i$ for some $k$: it is either empty or is the interval $[0,kq^i) = \bigcup_{0 \le l \le k} B_{i,l}$.

If the property is true for all $l$ with $0 \le l< j$, then $\Lambda^{-1}(\bigcup_{0 \le l < j} B_{i,l}) = \bigcup_{0 \le l \le k'} B_{i,l}$ for some $k'$. Again, since $|\Lambda^{-1}(B_{i,j})| =kq^i$ for some $k$, we have $\Lambda^{-1}(B_{i,j}) = \bigcup_{k' < l \le k+k'} B_{i,l}$.

We now prove the lemma itself. Suppose $a, b \in A^n$ satisfy $(a_i,\ldots,a_n) = (b_i,\ldots,b_n)$, then $a,b \in B_{i-1,j}$ where $j = \sum_{l=i}^n a_l$. By our claim, $\Lambda(B_{i-1,j})$ is an interval contained in a block $B_{i-1,k}$ for some $k$. Hence $\Lambda(a)_l = \Lambda(b)_l$ for all $l \ge i$.
\end{proof}

\begin{proposition} \label{prop:p_lambda}
Let $f$ be a permutation of $A^n$ which can be computed as a product of $n$ instructions updating $y_1$ to $y_n$. Then for any proper integer partition $\lambda$ of $q^n$, the transformation $g = f \circ \Lambda$ can also be computed as a product of $n$ instructions updating $y_1$ to $y_n$.
\end{proposition}

\begin{proof}
Let $f = f^n \circ \cdots \circ f^1$, where $f^i$ is an instruction updating $y_i$ for all $i$. Let $g^i$ be the transformation obtained after the instructions $y_m \gets g_m(y)$ for $m$ from $1$ to $i$; we have
$$
    g^i(x) = (g_1(x),\ldots,g_i(x),x_{i+1},\ldots,x_n).
$$
Then we only need to prove that for all $1 \le i \le n-1$ and all $a \ge b \in A^n$,
$g^i(a) = g^i(b) \Rightarrow g(a) = g(b).$

For any $m \le i$, we have $g^i_m = g_m = (f \circ \Lambda)_m = f_m \circ p$. Therefore, $g^i(a) = g^i(b)$ if and only if $f_m(\Lambda(a)) = f_m(\Lambda(b))$ for all $m \le i$ and $a_l = b_l$ for all $l \ge i+1$. By Lemma \ref{lemma:compatible_suffix}, we obtain $\Lambda(a)_l = \Lambda(b)_l$ for all $l \ge i+1$. Thus $g^i(a) = g^i(b)$ implies $h(\Lambda(a)) = h(\Lambda(b))$, where
$$
    h(x) = (f^i \circ \cdots \circ f^1) (x) = (f_1(x),\ldots,f_i(x),x_{i+1},\ldots,x_n).
$$
Since $h$ is a permutation, we obtain $\Lambda(a) = \Lambda(b)$ and hence $g(a) = g(b)$.
\end{proof}

\begin{theorem} \label{th:4n-3}
Any transformation of $A^n$ can be computed in at most $4n-3$ instructions.
\end{theorem}

\begin{proof}
Let $f$ be a transformation of $A^n$ and consider the integer partition $\mu$ of $q^n$ with $\mu_a = |f^{-1}(a)|$ for all $a \in A^n$. Sort $\mu$ properly: we obtain $\lambda_a = |f^{-1}(h(a))|$ for some permutation $h$ of $A^n$. Then $f$ can be expressed as
$f = h \circ \Lambda \circ g,$
where $g$ is a permutation of $A^n$ satisfying
$$
    g(f^{-1}(h(a))) = \left[\sum_{b=0}^{a-1} \lambda_b, \sum_{b=0}^a \lambda_b \right),
$$
for all $a \in A^n$.

By Theorem \ref{th:2n-1_permutation}, $g$ and $h$ can be computed as follows, where the superscript indicates which coordinate is updated by each instruction:
\begin{align*}
    g &=  \bar{g}^1\circ \cdots \circ \bar{g}^{n-1} \circ g^n \circ \cdots \circ g^1,\\
    h &=  \bar{h}^1\circ \cdots \circ \bar{h}^{n-1} \circ h^n \circ \cdots \circ h^1.
\end{align*}
By Proposition \ref{prop:p_lambda}, the transformation $h^n \circ \cdots \circ h^1 \circ \Lambda$ can be computed in $n$ instructions $\Lambda^n \circ \cdots \circ \Lambda^1$. Furthermore, $\Lambda^1$ and $\bar{g}^1$ being instructions updating $y_1$, their product $q^1 = \Lambda^1 \circ \bar{g}^1$ is another instruction updating $y_1$. Thus, $f$ can be computed by the following program of length $4n-3$:
$$
    f = \bar{h}^1\circ \cdots \circ \bar{h}^{n-1} \circ \Lambda^n \circ \cdots \circ \Lambda^2 \circ q^1 \circ \bar{g}^2 \circ \cdots \circ \bar{g}^{n-1} \circ g^n \circ \cdots \circ g^1.
$$
\end{proof}


We conclude this section with a remark on infinite alphabets. If $A$ is infinite, there exists a bijection $h : A^n \to A$ and
thus any transformation can be computed in $n+1$ instructions by the following program:
\begin{align*}
    y_n &\gets h(y)\\
    y_1 &\gets f_1(x)\\
    &\vdots\\
    y_n &\gets f_n(x).
\end{align*}
Therefore, considering a finite alphabet $A$ is not only interesting for applications, but it also leads to nontrivial effects. Note, however, that computing linear transformations of $\mathbb{Z}^n$ by linear instructions has been considered in \cite{AB09a}.

\section{Manipulating variables} \label{sec:manipulation}

We generalise the example of swapping two variables by considering any manipulation of variables. We distinguish between a transformation $\phi$ of $[n]$ (where we denote $[n] = \{1,\ldots,n\}$) which represents the formal movement of variables and the transformation $f^\phi$ of $A^n$ it induces on all the possible values of the variables. We remark that $f^\phi \in \Sym(A^n)$ if and only if $\phi \in \Sym(n)$. We always use the postfix notation for $\phi$, i.e. the image of $i$ under $\phi$ is denoted as $i \phi$. For $\phi : [n] \to [n]$, $\phi^k$ does represent the $k$-th power of $\phi$ according to composition.

\begin{definition}
A {\em manipulation of variables} is a transformation $f^\phi$ of $A^n$ of the form
$$
    f^\phi(x_1,\ldots,x_n) = (x_{1\phi}, \ldots, x_{n\phi})
$$
for some transformation $\phi$ of $[n]$.
\end{definition}

The transformation $\phi$ can be represented using a directed graph on $[n]$ with $n$ arcs $(i,i\phi)$ (see \cite{GM09} for a detailed review of this representation of transformations). This directed graph has cycles of two kinds:
\begin{itemize}
    \item A cycle $(i, i\phi,\ldots,i\phi^{k-1})$ (where $i\phi^k = i$) is {\em detached} if for all $0 \le l \le k-1$, there is no $j_l \ne i\phi^{l-1}$ such that $j_l\phi = i\phi^l$. Equivalently, the cycle is an entire connected component of the graph.

    \item A cycle $(i, i\phi,\ldots,i\phi^{k-1})$ is {\em attached} otherwise, i.e. if there exists $0 \le l \le k-1$ and $j \in [n], j \ne i\phi^{l-1}$ such that $j\phi = i\phi^l$.
\end{itemize}
Note that if $\phi$ is a permutation, then all its cycles are detached.

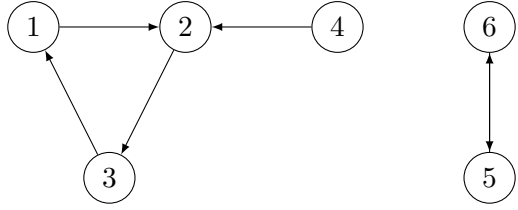
\begin{figure}
\begin{center}
\begin{tikzpicture}
    \node[draw, shape=circle] (1) at (0,2) {$1$};
    \node[draw, shape=circle] (2) at (2,2) {$2$};
    \node[draw, shape=circle] (3) at (1,0) {$3$};
    \node[draw, shape=circle] (4) at (4,2) {$4$};
    \node[draw, shape=circle] (5) at (6,0) {$5$};
    \node[draw, shape=circle] (6) at (6,2) {$6$};

    \draw[-latex] (1) -- (2);
    \draw[-latex] (2) -- (3);
    \draw[-latex] (3) -- (1);
    \draw[-latex] (4) -- (2);
    \draw[-latex] (5) -- (6);
    \draw[-latex] (6) -- (5);
\end{tikzpicture}
\caption{Representing a transformation via a graph.}
\label{fig:phi}
\end{center}
\end{figure}

For instance, consider $\phi : [6] \to [6]$ defined as $1\phi = 2$, $2\phi = 3$, $3\phi = 1$, $4\phi = 2$, $5\phi = 6$, $6\phi = 5$. Then the cycle $(1,2,3)$ is attached to $4$, while the cycle $(5,6)$ is detached, as seen on Figure~\ref{fig:phi}.

Let us first consider the case of a cyclic shift of variables. A similar result to Proposition \ref{prop:cycle} below is given in \cite{BGT13}.

\begin{proposition} \label{prop:cycle}
Let $\kappa \in \Sym(n)$ be a cyclic permutation, without loss $\kappa = (1, 2, \ldots, n)$. Then the cyclic shift of $n$ variables $f^\kappa: A^n \to A^n$ can be computed in $n+1$ instructions if and only if the order of updates (up to starting point) is $y_1,y_n,\ldots,y_2,y_1$.
\end{proposition}

\begin{proof}
Let us prove that if the order is correct, then we can compute the cyclic shift. This is done via the following program:
\begin{align*}
    y_1 &\gets \sum_{i=1}^n y_i\\
    y_n &\gets y_1 - \sum_{j=2}^n y_j\\
    &\vdots\\
    y_1 &\gets y_1 - \sum_{j=2}^n y_j.
\end{align*}
We prove the correctness of this program by induction: we claim that after the update of $y_{n-i}$, all variables $y_n,y_{n-1},\ldots,y_{n-i}$ have the correct values $x_1,x_n,\ldots,x_{n-i+1}$ for $i$ from $0$ to $n-1$. For $i=0$, we have
$$
    y_n \gets y_1 - \sum_{j=2}^n y_j = \sum_{i=1}^n x_i - \sum_{j=2}^n x_j = x_1.
$$
Now suppose it holds for up to $i-1$, we then have
$$
    y_{n-i} \gets y_1 - \sum_{j=2}^{n-i} y_j - \sum_{j=n-i+1}^n y_j = \sum_{i=1}^n x_i - \sum_{j=2}^{n-i} x_j - \sum_{k=n-i+2}^n x_k - x_1 = x_{n-i+1}.
$$

We now prove the reverse implication. Consider a program computing the shift of variables with $n+1$ instructions, and let $y_1$ be updated first. Then, suppose $y_i$ is updated before $y_{i+1}$. After $y_i \gets x_{i+1}$, the content of $(y_i,y_{i+1})$ is $(x_{i+1},x_{i+1})$ and the resulting transformation is not a permutation. Thus, for any $1 \le i \le n-1$, the update of $y_i$ must occur after that of $y_{i+1}$ and the only possible order of updates is $y_1,y_n,\ldots,y_1$.
\end{proof}

\begin{example} Let $\pi = (1, 2, 3)$ and $f^\pi : A^3 \to A^3$ such that $f^\pi(x_1,x_2,x_3) = (x_2,x_3,x_1)$. This can be computed via linear combinations:
\begin{align*}
    y_1 &\gets y_1+y_2+y_3 \qquad (=x_1+x_2+x_3)\\
    y_3 &\gets y_1-y_2-y_3 \qquad (=x_1)\\
    y_2 &\gets y_1-y_2-y_3 \qquad (=x_3)\\
    y_1 &\gets y_1-y_2-y_3 \qquad (=x_2).
\end{align*}
However, it is impossible to perform this cyclic shift in four instructions by first updating $y_1$ and then updating $y_2$ instead of $y_3$.
\end{example}

We can then determine the procedural complexity of a manipulation of variables.

\begin{theorem} \label{th:permutation}
Let $\phi: [n] \to [n]$ have $F$ fixed points and $D$ detached cycles. Then the procedural complexity of the manipulation of $n$ variables $f^\phi: A^n \to A^n$ is exactly
\begin{itemize}
    \item $n-F+D$ instructions if $\phi$ is a permutation;

    \item $n-F+1$ instructions if $\phi$ is not a permutation and $D > 0$;

    \item $n -F$ instructions otherwise.
\end{itemize}
\end{theorem}

\begin{proof}
Let us first suppose that $\phi$ is a permutation. Then computing one cycle after the other yields a program of length $n-F+D$ by Proposition \ref{prop:cycle}. Conversely, assume that there is a program computing $f^\phi$ in fewer than $n-F+D$ instructions. For this program there must be at least one cycle of $\phi$ such that each coordinate in the cycle is updated only once. Then after the first such update $y_i \gets x_{i\phi}$, we have $y_i = y_{i\phi} = x_{i\phi}$ and hence the resulting transformation is not a permutation.

Let us now suppose that $\phi$ is not a permutation. Let $m$ denote the number of variables which are not fixed and do not belong to any cycle. The subgraph induced on these vertices is acyclic, hence we can order them as $a_1,\ldots,a_m$ such that $a_i = a_j\phi$ only if $i > j$ \cite{BM08}. The first part of the program consists in updating all these vertices but the last in the correct order: for $i$ from $1$ to $m-1$, do
$$
    y_{a_i} \gets y_{a_i \phi}.
$$
The second part is to perform the cycles by using $y_{a_m}$ as memory. Let $\{i_c : 1 \le c \le C\}$ denote a member of each (detached or attached) cycle of length $l_c$, then do the following instruction:
$$
    y_{a_m} \gets \sum_{c=1}^C y_{i_c}.
$$
Then for all $c$ from $1$ to $C$ do
\begin{align*}
    y_{i_c} &\gets y_{i_c \phi}\\
    &\vdots\\
    y_{i_c \phi^{l_c - 2}} &\gets y_{i_c \phi^{l_c - 2}}\\
    y_{i_c \phi^{l_c-1}} &\gets y_{a_m} - \sum_{b=1}^{c-1} y_{i_b \phi^{l_b-1}} - \sum_{b=c+1}^C y_{i_b}.
\end{align*}
It can be easily proved by induction on $c$ that this program does compute all cycles. Eventually, we need the final update of $y_{a_m}$. Note that $a_m \phi$ is either a fixed point or it belongs to a cycle; therefore $x_{a_m \phi}$ is contained in $y_{a_m \phi^L}$, where $L=0$ if $a_m \phi$ is a fixed point and $L = l_c - 1$ if it belongs to the cycle $c$. Thus, the final update is given by
\begin{equation} \label{eq:y_a_m2}
    y_{a_m} \gets y_{a_m \phi^L}.
\end{equation}
Since $y_{a_m}$ is the only coordinate updated twice, this program has length $n-F+1$.

We now simplify this program when $\phi$ has no detached cycles. This time, for $i$ from $1$ to $m$, do
$$
    y_{a_i} \gets y_{a_i \phi}.
$$
Then for all $c$ from $1$ to $C$, there exists $\alpha_c \in \{a_1,\ldots,a_m\}$ such that $\alpha_c \phi = i_c$, therefore do
\begin{align*}
    y_{i_c} &\gets y_{i_c \phi}\\
    &\vdots\\
    y_{i_c \phi^{l_c - 2}} &\gets y_{i_c \phi^{l_c - 2}}\\
    y_{i_c \phi^{l_c-1}} &\gets y_{\alpha_c}.
\end{align*}
Since $y_{a_m}$ already contains $x_{a_m \phi}$, there is no need to include the final update in (\ref{eq:y_a_m2}).

Conversely, it is clear that at least $n-F$ instructions are needed to compute $f^\phi$. Furthermore, assume $D>0$ and that there is a program computing $f^\phi$ in exactly $n-F$ instructions. Let $i$ in the cycle $c$ be the first coordinate belonging to a detached cycle to be updated. Then the program first does $y_i \gets x_{i\phi}$ and the value of $x_i$ is lost; therefore, the update $y_{i \phi^{l_c - 1}} \gets x_i$ cannot occur.
\end{proof}

Theorem~\ref{th:permutation} indicates that disjoint cycles of a permutation cannot be computed ``concurrently,'' for the shortest program which computes two cycles exactly consists of computing one before the other.

\begin{corollary} \label{cor:transpositions}
If $n=2m$, then computing $m$ disjoint transpositions of variables (e.g. $(1, 2)$ $(3, 4)$ $\cdots$ $(2m-1,  2m)$) takes exactly $3m$ instructions. If $n=2m+1$, then computing $m-1$ disjoint transpositions and a cycle of length $3$, (e.g. $(1, 2)(3, 4)\cdots(2m-3, 2m-2)(2m-1,  2m, 2m+1)$) takes exactly $3m+1$ instructions. This is the maximum number of instructions for any manipulation of variables.
\end{corollary}

In particular, if $x_1,\ldots,x_{m^2}$ are the entries of an $m \times m$ matrix over $A$, then transposing that matrix takes exactly $3m(m-1)/2$ instructions.

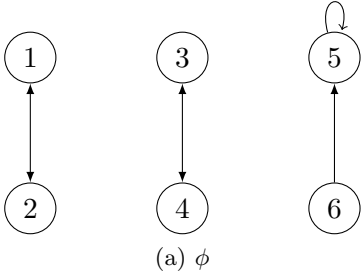
\begin{figure}
\centering
\subfloat[$\phi$]{\begin{tikzpicture}
    \node[draw, shape=circle] (1) at (0,2) {$1$};
    \node[draw, shape=circle] (2) at (0,0) {$2$};
    \node[draw, shape=circle] (3) at (2,2) {$3$};
    \node[draw, shape=circle] (4) at (2,0) {$4$};
    \node[draw, shape=circle] (5) at (4,2) {$5$};
    \node[draw, shape=circle] (6) at (4,0) {$6$};

    \draw[latex-latex] (1) -- (2);
    \draw[latex-latex] (3) -- (4);
    \draw[-latex] (6) -- (5);

    \path[-latex] (5) edge [loop above] node {} (5);
\end{tikzpicture}} \hfill
\subfloat[Programs for $f^\phi$]{\begin{tabular}[b]{|ll|ll|}
    \hline
    \multicolumn{2}{|c|}{With combinations} & \multicolumn{2}{|c|}{Black box}\\
    \hline
    $y_6 \gets y_1 + y_3$ &$(=x_1 + x_3)$ & $y_6 \gets y_1$ &$(=x_1)$\\
    $y_1 \gets y_2$  & $(=x_2)$ & $y_1 \gets y_2$ &$(=x_2)$\\
    $y_2 \gets y_6 - y_3$ &$(=x_1)$ & $y_2 \gets y_6$ &$(=x_1)$\\
    $y_3 \gets y_4$ &$(=x_4)$ & $y_6 \gets y_3$ &$(=x_3)$\\
    $y_4 \gets y_6 - y_2$ &$(=x_3)$ & $y_3 \gets y_4$ &$(=x_4)$\\
    $y_6 \gets y_5$ &$(=x_5)$ & $y_4 \gets y_6$ &$(=x_3)$\\
     & & $y_6 \gets y_5$ &$(=x_5)$\\
    \hline
\end{tabular}}
\caption{The simplest manipulation of variables using a shorter program with arithmetic} \label{fig:phi7}
\end{figure}

Another consequence of Theorem \ref{th:permutation} is that when $\phi$ is not a permutation, we can obtain shorter programs by using some arithmetic than by adopting the ``black box'' approach used for the swap of two variables described in the very beginning of the paper. Figure \ref{fig:phi7} shows the smallest example: computing $f^\phi$ takes $6$ instructions when using the program described in the proof of Theorem \ref{th:permutation}, while it takes $7$ instructions when we do not combine variables. Clearly, this example can be generalized by adding more cycles, thus yielding an arbitrarily large gap between the two approaches. The results are summarised in Proposition \ref{prop:black_box} and Corollary \ref{cor:phi}. We say an instruction is a {\em black box instruction} if it is of the form $y_i \gets y_j$ for $i,j \in [n]$.

\begin{proposition} \label{prop:black_box}
Let $\phi$ be a transformation of $[n]$ with $F$ fixed points and $D$ detached cycles. Then the manipulation of variables $f^\phi$ can be computed without memory by black box instructions if and only if $\phi$ is not a permutation (or is the identity). In that case, the shortest length of a black box program is $n-F+D$.
\end{proposition}

\begin{proof}
The proof calls arguments similar to those used above; as such, we use the same notation. We further enforce that the last $D-C$ elements $a_i$ are attached to different cycles, i.e. $a_{m-D+C+c}$ is attached to the cycle $c$. 

The following program computes $f^\phi$ in $n-F+D$ instructions. First, for $i$ from $1$ to $m-D+C$ do 
$$
	y_{a_i} \gets y_{a_i \phi}.
$$
Second, compute all detached cycles using $y_{a_m}$ as memory. For the detached cycle $\{i, i\phi, \ldots, i\phi^{l-1}\}$, do
\begin{align*}
	y_{a_m} &\gets y_i\\
	y_i &\gets y_{i \phi}\\
	&\vdots\\
	y_{i \phi^{l-2}} &\gets y_{i \phi^{l-1}}\\
	y_{i \phi^{l-1}} &\gets y_{a_m}.
\end{align*} 
This uses one extra instruction per detached cycle, i.e. $D$ extra instructions in total. Third, compute all the attached cycles, using $a_{m-D-C+c}$ as memory for the cycle $c$ (similar as above). This does not add any extra instruction.

It is clear that computing a detached cycle using instructions of the form $y_i \gets y_j$ requires using another variable as memory, and hence an extra instruction. However, since this variable gets a value from only one detached cycle, it cannot be re-used for the computation of any other detached cycle. Thus, we need at least $D$ extra instructions.
\end{proof}

It is worth noting that the proof of Proposition \ref{prop:black_box} does not use the fact that we are computing without memory. Therefore, the black-box computation will always take $n-F+D$ instructions, regardless of how much memory is used.

\begin{corollary} \label{cor:phi}
If $\phi$ is not a permutation, then the ratio between the procedural complexity of $f^\phi$ over the minimum length of a black box program computing $f^\phi$ is always greater than $2/3$. Conversely, for any $\epsilon > 0$, there exists $\phi$ for which that ratio is lower than $2/3+\epsilon$.
\end{corollary}

\begin{proof}
It takes at least $n-F$ instructions to compute $f^\phi$ without memory, and exactly $n-F+D$ instructions to do it using black box instructions. Since $n-F \ge 2D$, we easily obtain the lower bound of $2/3$.

Conversely, for any $k \ge 1$, let $n = 2k+2$ and $\phi: [n] \to [n]$ be defined as
$$
	\phi = (1,2) \circ \ldots \circ (2k-1,2k) \circ (2k+2 \to 2k+1).
$$
Then for any $A$, $f^\phi$ can be computed in $n$ instructions, but takes $3n/2-2$ instructions of the form $y_i \gets y_j$.
\end{proof}

\section{Using additional registers} \label{sec:memory}

In this section, we consider two different scenarios, which can be viewed as equivalent in the memoryless computation framework. The first scenario is when we have more registers that we need and hence we want to compute a function which only depends on and updates a limited number of registers. This is equivalent to computing a function of those registers and treating the remaining registers as memory cells which can be accessed as easily as the other registers. Thus, our second scenario (which we shall consider here) is when we want to compute a transformation $f$ of $A^n$ using $m$ memory cells storing values in $A$. 

By convention, we shall denote the content of the $m$ memory cells as $y_{n+1}, \ldots, y_{n+m}$; we still use $y = (y_1,\ldots,y_n)$. Then computing $f$ using $m$ memory cells is equivalent to computing some transformation $h(x_1,\ldots,x_{n+m})$ of $A^{n+m}$ such that the first $n$ coordinate functions of $h$ coincide with those of $f$. Let us denote the set of such transformations as $D(f,m)$. The shortest length of a program computing $f$ using $m$ memory cells is hence given by
$$
    \mathcal{L}(f|m) := \min_{h \in D(f,m)} \mathcal{L}(h).
$$
Therefore, there exists $h$ such that $\mathcal{L}(h) = \mathcal{L}(f|m)$ but it may be difficult to characterise that transformation $h$. However, Proposition~\ref{prop:length_memory} shows that there is a deterministically (and easily) described transformation $h \in D(f,m)$ for which $\mathcal{L}(h)$ and $\mathcal{L}(f|m)$ are in bijection. Therefore, the memoryless computation framework also considers the case of using memory.

\begin{proposition} \label{prop:length_memory}
For any transformation $f$ of $A^n$ and any $e = (e_1,\ldots,e_m) \in A^m$, let $h^e \in D(f,m)$ and $h^e_{n+i} = e_i$ for $1 \le i \le m$. Then
$$
    \mathcal{L}(h^e) = \mathcal{L}(f|m) + m.
$$
\end{proposition}

\begin{proof}
Let $g \in D(f,m)$ such that $\mathcal{L}(g) = \mathcal{L}(f|m)$, then the shortest program computing $g$ appended with the suffix $y_{n+i} \gets e_i$ for $i$ from $1$ to $m$ has length $\mathcal{L}(f|m) + m$ and computes $h^e$. Therefore, $\mathcal{L}(h^e) \le \mathcal{L}(f|m) + m$.

Conversely, consider the shortest program computing $h^e$. It contains $m$ final updates $y_{n+i} \gets e_i$ which, without loss, appear for $i$ from $m$ down to $1$. Then any instruction $y_j \gets g(y)$ occurring after $y_{n+k} \gets e_k$ (hence $j \le n+k-1$) can be replaced by $y_j \gets g'(y_1,\ldots,y_{n+k-1})$ where $g' : A^{n+k-1} \to A$ is defined as
$$
    g'(y_1,\ldots,y_{n+k-1}) = g(y_1,\ldots,y_{n+k-1},e_k,\ldots,e_m).
$$
Now remove all the $y_{n+i} \gets e_i$ updates; we are left with a program computing some transformation in $D(f,m)$ of length $\mathcal{L}(h^e) - m$. Thus $\mathcal{L}(f|m) \le \mathcal{L}(h^e) - m$.
\end{proof}

There is a linear analogue to Proposition \ref{prop:length_memory}. Namely, if $M \in \GF(q)^{n \times n}$, let $N \in \GF(q)^{n+m \times n} = (N,0)$. Then it is easily shown that, when limiting ourselves to linear instructions, the procedural complexity of $N$ is equal to $m$ plus the minimum length of a program computing $M$ with $m$ memory cells. 

\subsection{Shorter programs}

We have shown in Theorem~\ref{th:all} that one need not use memory to compute any transformation. However, we shall prove that one may want to use memory in order to use shorter programs. 

We have shown in Theorem~\ref{th:2n-1_permutation} that any permutation can be computed without memory in at most $2n-1$ instructions. On the other hand, using one memory cell necessarily yields a program with length at least $n+1$. Propositions~\ref{prop:ab} and~\ref{prop:ab_1} show that these two results are simultaneously tight: there exists a permutation $f \in \Sym(A^n)$ for which $\mathcal{L}(f) = 2n-1$ while $\mathcal{L}(f|1) = n+1$.

\begin{proposition} \label{prop:ab_1}
The transposition $(a, b)$ of two states $a,b \in A^n$ at Hamming distance $d$ can be computed with one memory cell in $d+1$ instructions: $\mathcal{L}((a, b)|1) = d+1$.
\end{proposition}

\begin{proof}
Without loss, let $a$ and $b$ disagree on their first $d$ coordinates. Then the following program computes $(a, b)$:
\begin{align*}
    y_{n+1} &\gets \delta(y,a) - \delta(y,b)\\
    y_1 &\gets y_1 + (b_1 - a_1)y_{n+1}\\
    &\vdots\\
    y_d &\gets y_d + (b_d - a_d)y_{n+1}.
\end{align*}
\end{proof}

In Theorem \ref{th:4n-3}, we have given an upper bound on the complexity of any transformation which only depends on the number of variables. This upper bound is larger than $2n-1$ obtained for permutations; however, using memory cells yields a program using $2n-1$ instructions, as seen below.

\begin{proposition} \label{prop:2n-1}
Any transformation $f$ of $A^n$ can be computed with $n-1$ memory cells and no more than $2n-1$ instructions: $\mathcal{L}(f|n-1) \le 2n-1$.
\end{proposition}

\begin{proof}
The following program computes $f$ using $n-1$ memory cells and $2n-1$ instructions:
\begin{align*}
    y_{n+1} &\gets y_1\\
    &\vdots\\
    y_{2n-1} &\gets y_{n-1}\\
    y_1 &\gets f_1(y_{n+1},\ldots,y_{2n-1},y_n)\\
    &\vdots\\
    y_n &\gets f_n(y_{n+1},\ldots,y_{2n-1},y_n).
\end{align*}
\end{proof}

Proposition~\ref{prop:2n-1} indicates that we do not need any more than $n-1$ memory cells. Indeed, if we use $n$ memory cells, then the program will have at least $2n$ instructions (unless some memory cells are not updated, which is equivalent to not using them). Therefore, $\mathcal{L}(f|m) = \mathcal{L}(f|n-1)$ for any $m \ge n-1$.

We remark that this upper bound on the amount of memory needed follows from the fact that we allow any instruction. In practice, using a large amount of memory is the price paid for using only a restricted number of basic instructions.


The ideas behind Theorem~\ref{th:2n-1_permutation} can be adapted to the case of using memory to yield a refinement of Proposition~\ref{prop:2n-1} for permutations.

\begin{theorem} \label{th:n/2_memory}
Any permutation of $A^n$ can be computed in at most $3m$ instructions with $m$ memory cells if $n=2m$ is even and at most $3m+3$ instructions with $m+2$ memory cells if $n=2m+1$ is odd.
\end{theorem}

\begin{proof}
Suppose $n=2m$ and let $f \in \Sym(A^n)$. By Proposition~\ref{prop:exchange}, there exist $m$ functions $g_1,\ldots,g_m : A^n \to A$ such that
$$(f_1,\ldots,f_m,g_1,\ldots,g_m) \qquad \mbox{and} \qquad (x_{m+1},\ldots,x_n,g_1,\ldots,g_m)$$ both form permutations of $A^n$. The program goes as follows:
\begin{itemize}
    \item Step 1 ($m$ instructions). For $i$ from $1$ to $m$, do $y_{n+i} \gets g_i(x)$.

    \item Step 2 ($m$ instructions). For $i$ from $1$ to $m$, do $y_i \gets f_i(x)$. This is possible since 
	$$(x_{m+1},\ldots,x_n,g_1,\ldots,g_m)$$
	form a permutation of $A^n$, and hence $f_i(x)$ can be expressed as a function of $$(y_{m+1},\ldots,y_n,y_{n+1},\ldots,y_{n+m}).$$

    \item Step 3 ($m$ instructions). For $i$ from $m+1$ to $n$, do $y_i \gets f_i(x)$. This is possible since $(f_1,\ldots,f_m,g_1,\ldots,g_m)$ form a permutation of $A^n$, and hence $f_i(x)$ can be expressed as a function of $(y_1,\ldots,y_m,y_{n+1},\ldots,y_{n+m})$.
\end{itemize}

Now let $n=2m+1$ be odd. Then add one memory cell and consider the extended permutation $g \in D(f,1)$ such that $g_{2m+2}(x) = x_{2m+2}$. Then $g$ can be computed in $3m+3$ instructions and $m+1$ memory cells.
\end{proof}

Therefore, we do not want more than around $n/2$ memory cells to compute any permutation; adding any more would be superfluous. There is a linear analogue to Theorem~\ref{th:n/2_memory}.

\begin{proposition}
Any linear permutation of $A^n$ can be computed in at most $3m$ linear instructions with $m$ memory cells if $n=2m$ is even and at most $3m+3$ linear instructions with $m+2$ memory cells if $n=2m+1$ is odd.
\end{proposition}

\begin{proof}
Suppose $n=2m$. Let $f(x) = xM^\top$ and denote the first $m$ rows of $M$ as $M_1$ and the matrix $J = (0|I_m) \in A^{m \times n}$. We claim that there exists a matrix $N \in A^{m \times n}$ such that $(M_1^\top,N^\top)$ and $(J^\top,N^\top)$, both in $A^{n \times n}$, are nonsingular. Then the algorithm simply places $N$ in the memory, then replaces the first $m$ rows by $M_1$, and finally updates the last $m$ rows to those of $M$.

We now justify our claim. This is equivalent to showing that for any two subspaces in the Grassmannian $G(q,2m,m)$ of $m$-dimensional subspaces of $\GF(q)^{2m}$, there exists a third subspace in the same Grassmannian at subspace distance $2m$ from both \cite{KK08} (where the subspace distance between $U,V \in G(q,2m,m)$ is given by $2 \dim(U + V) - 2m$). Since the Grassmannian endowed with the subspace distance forms an association scheme \cite{Del76}, we only have to check for the row space of $J$ and one subspace at distance $2d$ for each $0 \le d \le m$. Let us then assume $M_1 = (0_{m-d} | I_m | 0_d)$ whose row space is at subspace distance $2d$ from that of $J$. Then it is easily checked that the row space of
$$
    N = \left( \begin{array}{c|c|c}
    \multirow{2}{*}{$I_m$} & \multirow{2}{*}{$0_d$} & 0_{m-d} \\
    \cline{3-3}
    & & I_{m-d}
    \end{array}\right)
$$
is at distance $2m$ from the row spaces of $M_1$ and $J$.

The case $n=2m+1$ is settled by considering $M' \in A^{n+1 \times n+1}$ given by
$$
    M' = \left(\begin{array}{c|c}
    M & 0\\
    \hline
    0 & 1
    \end{array}\right).
$$
\end{proof}

For manipulations of variables, we can completely determine the gain offered by using memory. In particular, using only one memory cell is optimal to compute any manipulation of variables.

\begin{proposition} \label{prop:permutation_one_memory}
Any manipulation of $n$ variables with $F$ fixed points can be computed with one memory cell in at most $n-F+1$ instructions.
\end{proposition}

\begin{proof}
By Theorem \ref{th:permutation}, we only need to prove the case where $\phi$ is a permutation of $[n]$. Let $\pi$ be the transformation of $[n+1]$ defined as $i\pi = i\phi$ for all $i \in [n]$ and $(n+1)\pi = 1$. Then by Theorem \ref{th:permutation}, we can compute $f^\pi$ in $n-F+2$ instructions, where the last instruction updates $y_{n+1}$. By removing that last instruction, we compute $f^\phi$ in $n-F+1$ instructions while using one memory cell $y_{n+1}$.
\end{proof}

By comparing with Theorem~\ref{th:permutation}, we see that using only one memory cell reduces the length of the program from $n-F+D$ to $n-F+1$ for permutations. In particular, for a disjoint product of $m$ transpositions, the complexity goes down from $3m$ to only $2m+1$.

\begin{example} Let $\pi = (1, 2)(3, 4) \in \Sym(4)$ and let $f^\pi : A^4 \to A^4$ be the corresponding permutation of variables. By Corollary~\ref{cor:transpositions}, two disjoint transpositions of variables must be computed in at least $6$ instructions when no memory is used. However, adjoining one memory cell $y_5$ leads to a program with only $5$ instructions, as seen below.
\begin{center}
\begin{tabular}{rcll}
    $y_5$ &$\gets$& $y_1 + y_3$ & $(= x_1 + x_3)$\\
    $y_1$ &$\gets$& $y_2$ & $(=x_2)$\\
    $y_2$ &$\gets$& $y_5 - y_3$ & $(=x_1)$\\
    $y_3$ &$\gets$& $y_4$ & $(=x_4)$\\
    $y_4$ &$\gets$& $y_5 - y_2$ & $(=x_3)$
\end{tabular}
\end{center}
\end{example}

\subsection{Binary instructions}

Since the number of instructions is very large, one may want to use only a subset of instructions to compute any transformation. A natural choice is that of binary instructions, since any function can be computed as a composition of binary operations.

\begin{definition} \label{def:binary}
An instruction $y_i \gets g_i(y)$ is {\em binary} if $g$ only involves at most two variables: $g_i(y) = g_i(y_j,y_k)$ for some $j,k \in [n]$.
\end{definition}


Using binary instructions is not sufficient when computing without memory; however, it is sufficient when only one memory cell is used.

\begin{theorem} \label{th:binary_instructions}
If $A = \GF(2)$, then the set of all permutations of $A^n$ which can be computed using binary instructions is the affine group $\Aff(n,2)$. On the other hand, when using one memory cell, any transformation over any alphabet can be computed by binary instructions.
\end{theorem}

\begin{proof}
Note that any binary permutation instruction is of the form $y_i \gets g_i(y_i,y_j)$ for some $j \in [n]$. If $A = \GF(2)$ and $n=2$, then it is well known that $\Sym(\GF(2)^2) = \Aff(2,2)$. If $n > 2$, then any instruction of the form $y_i \gets g(y_i,y_j)$ must correspond to a binary instruction for $\GF(2)^2$ acting on the coordinates $y_i$, $y_j$: it is also affine. Therefore, the group generated by binary permutation instructions is affine. Conversely, extending Gaussian elimination to the affine case shows that any affine permutation can be computed via binary instructions.

If one memory cell is used, we claim that the instructions in Theorem \ref{th:all} can be computed by binary instructions. For the sake of simplicity, let us assume $i=1$. For any $u \in A^n$ and $v = u + e^1$, we can decompose
\begin{align*}
    \delta(y,u) &= \delta(y_1,u_1) \delta(y_2,u_2) \cdots \delta(y_n,u_n),\\
    \delta(y,u) - \delta(y,v) &= (\delta(y_1,u_1) - \delta(y_1,v_1)) \delta(y_2,u_2) \cdots \delta(y_n,u_n).
\end{align*}
Then the transposition $(u, v)$ is computed as follows:
\begin{align*}
    y_{n+1} &\gets \delta(y_1,u_1) - \delta(y_1,v_1)\\
    y_{n+1} &\gets y_{n+1} \delta(y_2,u_2)\\
    &\vdots\\
    y_{n+1} &\gets y_{n+1} \delta(y_n,u_n)\\
    y_1 &\gets y_1 + y_{n+1}
\end{align*}
and the assignment $(e^0 \to e^1)$ is computed as:
\begin{align*}
    y_{n+1} &\gets \delta(y_1,0)\\
    y_{n+1} &\gets y_{n+1} \delta(y_2,0)\\
    &\vdots\\
    y_{n+1} &\gets y_{n+1} \delta(y_n,0)\\
    y_1 &\gets y_1 + y_{n+1}.
\end{align*}
Since any transformation can be computed using these two types of instructions, it can be computed with binary instructions.
\end{proof}

\section{Acknowledgment}

The authors would like to thank Peter J. Cameron, Ben Fairbairn, Peter Keevash and Rasmus Petersen for stimulating discussions.

\bibliographystyle{elsarticle-num}
\bibliography{g}

\begin{thebibliography}{10}
\expandafter\ifx\csname url\endcsname\relax
  \def\url#1{\texttt{#1}}\fi
\expandafter\ifx\csname urlprefix\endcsname\relax\def\urlprefix{URL }\fi
\expandafter\ifx\csname href\endcsname\relax
  \def\href#1#2{#2} \def\path#1{#1}\fi

\bibitem{Bur96}
S.~Burckel, Closed iterative calculus, Theoretical Computer Science 158 (1996)
  371--378.

\bibitem{BGT13}
S.~Burckel, E.~Gioan, E.~Thom\'e, Computation with no memory, and rearrangeable
  multicast networks, submitted.~Available at
  \url{http://arxiv.org/abs/1310.5380}.

\bibitem{YLCZ06}
R.~W. Yeung, S.-Y.~R. Li, N.~Cai, Z.~Zhang, Network Coding Theory, Vol.~2 of
  Foundation and Trends in Communications and Information Theory, now
  Publishers, Hanover, MA, 2006.

\bibitem{ACLY00}
R.~Ahlswede, N.~Cai, S.-Y.~R. Li, R.~W. Yeung, Network information flow, IEEE
  Transactions on Information Theory 46~(4) (2000) 1204--1216.

\bibitem{HP11}
J.~Hennessy, D.~Patterson, Computer Architecture: a quantitative approach, 5th
  Edition, Morgan Kaufmann, 2011.

\bibitem{Bur04}
S.~Burckel, Elementary decompositions of arbitrary maps over finite sets,
  Journal of Symbolic Computation 37~(3) (2004) 305--310.

\bibitem{BM00}
S.~Burckel, M.~Morillon, Three generators for minimal writing-space
  computations, Theoretical Informatics and Applications 34 (2000) 131--138.

\bibitem{BM04a}
S.~Burckel, M.~Morillon, Quadratic sequential computations of boolean mappings,
  Theory of Computing Systems 37~(4) (2004) 519--525.

\bibitem{BM04}
S.~Burckel, M.~Morillon, Sequential computation of linear boolean mappings,
  Theoretical Computer Science 314 (2004) 287--292.

\bibitem{BGT09}
S.~Burckel, E.~Gioan, E.~Thom\'e, Mapping computation with no memory, in: Proc.
  International Conference on Unconventional Computation, Ponta Delgada,
  Portugal, 2009, pp. 85--97.

\bibitem{Gua98}
D.-J. Guan, Generalized {G}ray codes with applications, Proc. Natl. Sci. Counc.
  ROC(A) 22~(6) (1998) 841--848.

\bibitem{GM09}
O.~Ganyushkin, V.~Mazorchuk, Classical Finite Transformation Semigroups: An
  Introduction, Vol.~9 of Algebra and Applications, Springer-Verlag, London,
  2009.

\bibitem{CGR13}
P.~J. Cameron, M.~Gadouleau, S.~Riis, Combinatorial representations, Journal of
  Combinatorial Theory, Series A 120~(3) (2013) 671--682.

\bibitem{BM08}
J.~Bondy, U.~Murty, Graph Theory, Vol. 244 of Graduate Texts in Mathematics,
  Springer, 2008.

\bibitem{GR01}
C.~D. Godsil, G.~Royle, Algebraic Graph Theory, Vol. 207 of Graduate Texts in
  Mathematics, Springer-Verlag, 2001.

\bibitem{EGZ61}
P.~Erd\"os, A.~Ginzburg, A.~Ziv, Theorem in the additive number theory,
  Bulletin of the Research Council of Israel 10F (1961) 41--43.

\bibitem{AB09a}
M.~Ahmad, S.~Burckel, Sequential decompositions of operations and compiler
  optimization, Tech. rep., INRIA (2009).

\bibitem{KK08}
R.~K\"otter, F.~R. Kschischang, Coding for errors and erasures in random
  network coding, IEEE Transactions on Information Theory 54~(8) (2008)
  3579--3591.

\bibitem{Del76}
P.~Delsarte, Association schemes and $t$-designs in regular semilattices,
  Journal of Combinatorial Theory A 20~(2) (1976) 230--243.

\end{thebibliography}

\end{document}